\documentclass[aps,pra,reprint,superscriptaddress,showpacs,floatfix, longbibliography, 10pt]{revtex4-2}

\usepackage{xcolor}
\usepackage{braket}       
\usepackage{graphicx}      
\usepackage{longtable}     
\usepackage{xurl}          
\usepackage{bm}            
\usepackage{soul}
\usepackage{amsmath}
\usepackage{amssymb}
\usepackage{siunitx}
\usepackage{hyperref}
\usepackage{cleveref}
\hypersetup{
    colorlinks=true,    
    linkcolor=blue,     
    citecolor=blue,     
    filecolor=blue,     
    urlcolor=blue       
}

\begin{document}
\title{Improved delta-kick cooling with multiple non-ideal kicks}
\author{Harshil Neeraj}
\email          {harshil.neeraj@mail.utoronto.ca}
\email          {harshilneeraj@gmail.com}
\affiliation{Department of Physics, University of Toronto, 60 St. George Street, Toronto ON, M5S 1A7, Canada}

\author{David C. Spierings}
\affiliation{
   MIT-Harvard Center for Ultracold Atoms and Research Laboratory of Electronics, Massachusetts Institute of Technology, Cambridge, MA 02139, USA}

\author{Joseph McGowan IV} 
\affiliation{Department of Physics, University of Toronto, 60 St. George Street, Toronto ON, M5S 1A7, Canada}

\author{Nicholas Mantella}
\affiliation{Department of Physics, University of Toronto, 60 St. George Street, Toronto ON, M5S 1A7, Canada}

\author{Aephraim M. Steinberg}
\email{steinberg@physics.utoronto.ca}
\affiliation{Department of Physics, University of Toronto, 60 St. George Street, Toronto ON, M5S 1A7, Canada}

\begin{abstract}

Delta-kick cooling is a technique employed to achieve low kinetic temperatures by decreasing momentum width at the cost of increased position width. In an ideal implementation, this method uses a harmonic potential to deliver a single near-instantaneous momentum kick. In practice, potentials that are approximately harmonic near their center are commonly used. As a result, the breakdown of the harmonic approximation far from the center limits the cooling performance. Inspired by aberration cancellation in optics, we propose to use compound matter-wave lens systems for $\delta-$kick cooling with Gaussian potentials. By strategically combining attractive and repulsive kicks, we show that it is possible to mimic the effect of a harmonic potential. For a test case with reasonable experimental parameters, our method suggests a reduction in kinetic temperature by a factor of $2.5$ using a 2-pulse sequence and by a factor of $3.2$ using a 3-pulse sequence.

\end{abstract}

\maketitle

\section{Introduction}
Delta-kick cooling (DKC), or matter-wave lensing, is a common technique in ultra-cold atom experiments to reduce a sample's kinetic temperature  \citep{Chu:86, DKC97, ManipulationOfMotionalQuantumStatesOfNeutralAtoms, DKC_longitudinal_focussing, DKC2000AMS, DKC_using_Ioffe_Pritchard_trap, DKC_2008, 38pK_DKC, 50pK_Kasevich, matter_wave_interferometry_in_microgravity}. In DKC, a ``lensing'' potential is briefly applied to a freely expanding cloud of atoms with the goal of narrowing the cloud's momentum distribution. This works in analogy to a lens placed a focal length away from a point source of light in order to create a collimated beam \citep{DKC_longitudinal_focussing}. Variants of matter-wave lensing also involve abruptly reducing the trap frequency, causing the width of the cloud to oscillate; the shallow trap is eventually turned off when atoms reach the turning points of the potential, bringing them to rest \citep{Chu:86}. DKC works by reducing the momentum uncertainty at the expense of an increase in position uncertainty, resulting in cooling despite involving no dissipation and thereby preserving phase space density \footnote{Since the DKC process is unitary and does not provide cooling in a thermodynamic sense, the technique is also often referred to as delta-kick collimation, emphasizing its analogy with optical beam collimation.
}. DKC has been used to achieve record low temperatures \citep{38pK_DKC,50pK_Kasevich, space_DKC_picoKelvin} in the \si{\pico\kelvin} regime and is often the final stage of cooling in atom interferometry \citep{Why_momentum_width_matters_for_atom_interferometry_with_Bragg_pulses, matter_wave_interferometry_in_microgravity} or gravimetry \citep{DKC_gravimeter} experiments, where large coherence lengths are needed. DKC is a shortcut to adiabatic evolution, as the final state is the same as that achieved by slowly changing the trap frequency to always stay in the instantaneous eigenstate. In fact, a generalized DKC protocol is time-optimal for achieving a final harmonic oscillator eigenstate for a broad class of atomic systems \citep{DKC2000AMS, DKC_David2021}. 

\begin{figure*}
\includegraphics[width=1\textwidth]{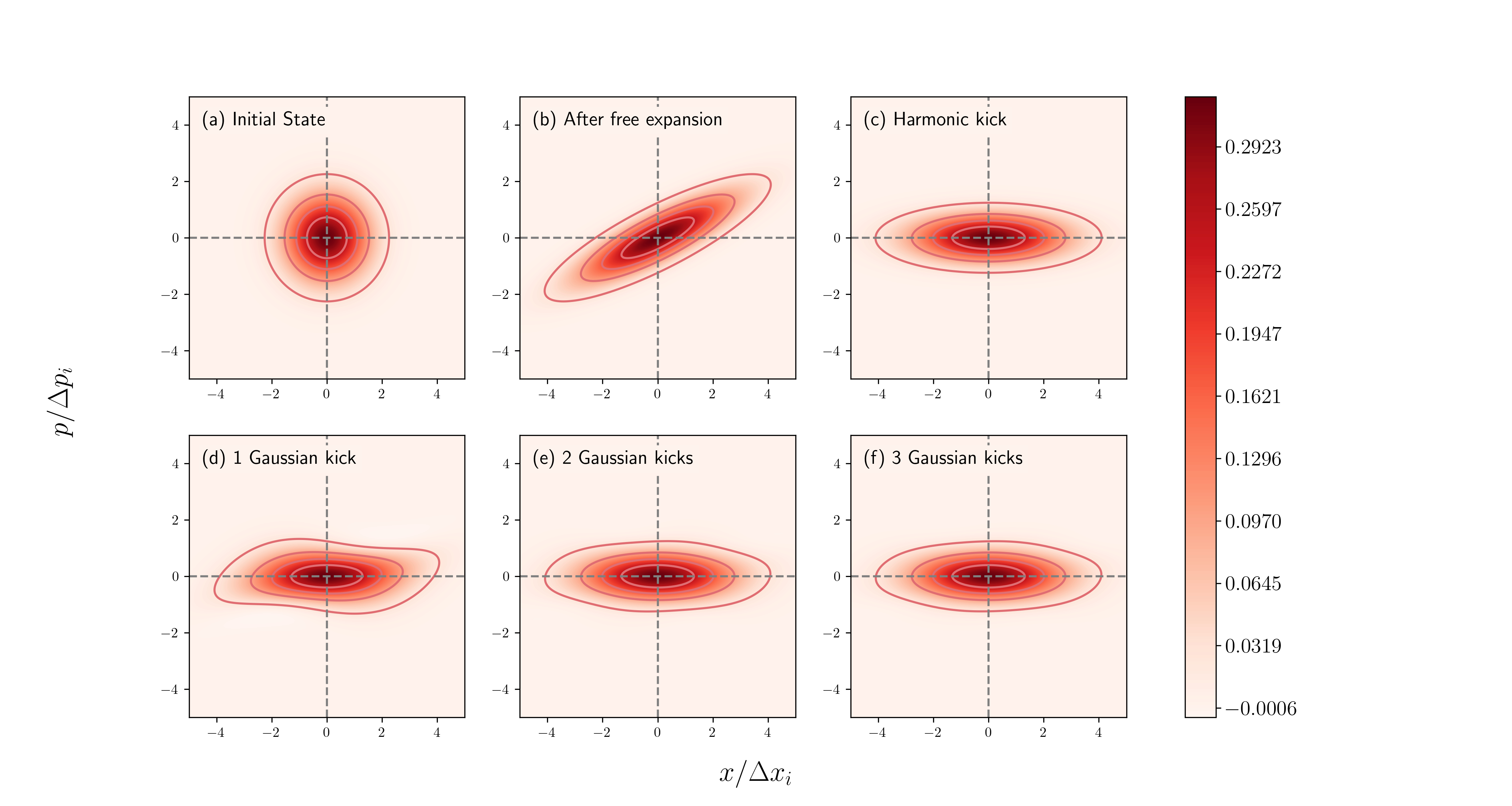}
    \caption{Wigner function during the standard DKC protocol (a-c) and after the proposed multi-kick scheme (d-f). (a) Atoms are prepared in the ground state of a harmonic trap of frequency $\omega_0$, with initial size $\Delta x_i = \sqrt{\hbar/2m\omega_0}$ and momentum width $\Delta p_i = \sqrt{\hbar m\omega_0/2}$. (b) After free expansion for time $t_f = 3/2\omega_0$, initial velocity and final position become correlated. A $\delta$-kick reduces the momentum width with (c) a harmonic potential, (d) a single Gaussian potential with RMS width $\sigma_1 = 5\Delta x_i$, (e) 2 Gaussian potentials - one attractive with $\sigma_1 = 5 \Delta x_i$ and one repulsive with $\sigma_2 = 4\Delta x_i$ and (f) 3 Gaussian potentials - attractive kicks of widths $\sigma_1 = 5 \Delta x_i, \sigma_3 = 3\Delta x_i$ and repulsive kick of width $\sigma_2 = 4\Delta x_i$.
}
\label{Wigner_fn_initial_free_harmonic_1_2_3_Gaussian}
\end{figure*}

DKC requires a harmonic potential to apply the correct momentum kick at all positions in space. We consider a non-interacting cloud of atoms that undergoes free expansion, followed by a harmonic potential flashed on for a short duration in time. In this ideal case, the ratio of the final and initial kinetic temperatures is given by 
\begin{equation}\label{eqn: T_f/T_i}
T_f/T_i \equiv  \left(\Delta v_f/\Delta v_i\right)^2 = \left(\Delta x_i/\Delta x_f\right)^2,    
\end{equation}
where $T = m (\Delta v)^2 /k_B$ for mass $m$ and Boltzmann constant $k_B$, and $\Delta v$ and $\Delta x$ quantify the RMS velocity and position widths at the initial and final times. The second equality in Eq.~(\ref{eqn: T_f/T_i}) follows from the conservation of phase space density. Fig.~\ref{Wigner_fn_initial_free_harmonic_1_2_3_Gaussian} (a)-(c) illustrates this ideal protocol. Free expansion stretches the phase space distribution along the position axis while conserving momentum width. A harmonic potential is then pulsed on, rotating the distribution to align its minimum uncertainty axis with the momentum axis.

A common implementation of DKC in cold atom experiments involves using the dipole force from a red-detuned Gaussian laser beam to apply the lensing potential. The approximately harmonic nature of the Gaussian potential near its center is used to apply a momentum kick that is roughly linear in position. For large clouds or long expansion times, however, the contribution from non-harmonic terms becomes significant, and the $\delta-$kick is not correct to bring all atoms to rest (Fig.~\ref{Wigner_fn_initial_free_harmonic_1_2_3_Gaussian} (d)). The beam size therefore often limits the lowest achievable temperature. 

Deviation of the lens potential from a quadratic can be considered analogous to non-parabolic lenses in optics, which give rise to spherical aberrations. Aberrations in DKC and in matter-wave optics have been discussed in \cite{50pK_Kasevich, multipole_representation_for_matter_wave_optics, Aberrations_in_DKC_2004}. ``Painting'' time-averaged optical potentials by modulating the trap center has been used to generate arbitrary potentials \citep{Time_avg_potential_expt_demonstration, DKC_time_avg_potential, DKC_time_avg_potential_2024_picoKelvin}, such as wide harmonic potentials for matter-wave lensing. Double-impulse magnetic lenses have been proposed for 3D focusing and aberration reduction in a cloud of launched cold atoms \cite{Two_magnetic_lens_2006}. In this work, we present a multiple kick approach to push the limits to DKC set by the finite beam size.

It is well known that aberrations in optics can be corrected by using a carefully chosen combination of converging and diverging lenses. In analogy with this, we propose to reduce aberrations in matter-wave lensing by using a combination of attractive and repulsive $\delta-$kicks acting as converging and diverging matter-wave lenses. We show that this increases the region of space where atoms experience an effective harmonic potential, allowing more initial free expansion and thereby improved cooling performance.

The remainder of the paper is structured as follows: In Sec.~\ref{Sec 2}, we review the typical frameworks for determining the kick strength of the matter-wave lens used in the ideal construction of DKC. In Sec.~\ref{Sec 3}, we suggest a method for balancing the ``aberrations'' in a compound matter-wave lens approach and determine the appropriate kick strengths of each lens. We evaluate the performance of these $\delta-$kicks quantum mechanically and compare them with that of a compound matter-wave lens system whose kick strengths are numerically optimized. 
We also evaluate the performance of a doublet matter-wave lens for different relative sizes of the two $\delta-$kick beams, as well as the sensitivity of the final velocity width on the exact value of kick strengths chosen.
Lastly, in Sec.~\ref{Sec 4}, we summarize our results.

\section{Aberrations in matter-wave lensing}\label{Sec 2}

A simple classical picture covers important features of DKC for a non-interacting atomic cloud. An atom with initial position $x_i$ and momentum $p_i$ will reach a final position \(x_f = x_i + p_i t_f/m \approx p_i t_f/m\) after a long free expansion time \(t_f\). A force that is linear in position, arising from a harmonic potential, can therefore provide an impulse that brings all the atoms to rest. The change in momentum of an atom is given by 
\begin{equation}\label{eqn: change in momentum for harmonic kick}
    \Delta p = - \delta t\left.\frac{d U}{d x}\right|_{x=x_f} = - m\omega_k^2x_f\delta t,
\end{equation}
where $U(x) = \frac{1}{2}m\omega_k^2x^2$ is the kick potential. The final momentum of the atom at position $x_f$ after the $\delta-$kick is then given by $p_f = p_i + \Delta p \approx m x_f(1/t_f - \omega_k^2  \delta t$). Hence, atoms can be brought to rest by applying a momentum kick such that $\omega_k^2\delta t = 1/t_f$. In a quantum mechanical calculation for atoms released from a 1D harmonic trap with trap frequency $\omega_0$, and expanding freely for a time $t_f$, the ideal $\delta-$kick is given by \cite{DKC_David2021} 
\begin{equation}\label{kick strength for harmonic kick}
\omega_k^2\delta t = \frac{\omega_0^2 t_f}{1+\omega_0^2 t_f^2}.
\end{equation} 
Eq.~(\ref{kick strength for harmonic kick}) reduces to the expression derived using classical analysis for large expansion times $\left(t_f\gg \Delta x_i/\Delta v_i = 1/\omega_0\right)$. The expansion time \(t_f\) is also referred to as the focal time in analogy to the focal length of a lens. Longer focal times imply a bigger position width after free expansion, translating to a smaller momentum width after the $\delta-$kick. 

A harmonic approximation is only valid near the trap center for most potentials. For a $\delta-$kick involving a Gaussian potential of the form $U(x) = U (1-e^{-x^2/2\sigma^2})$, the final momentum after the kick is given by
\begin{multline}\label{eqn_momentum with single Gaussian kick}
 p_f \approx \frac{m}{t_f}x_f- \frac{U \delta t}{\sigma^2}x_f e^{-x_f^2/2\sigma^2} \\
    \approx x_f\left(\frac{m}{t_f} - \frac{U\delta t}{\sigma^2}\right) +\frac{U\delta t}{2\sigma^4}x_f^3- \frac{U\delta t}{8\sigma^6}x_f^5+\hdots
\end{multline}
Unlike a harmonic kick, a single Gaussian kick cannot make the final momentum zero for atoms at all positions $x_f$. The non-harmonic terms in the potential result in ``aberrations'' in matter-wave lensing. For a cloud of atoms, this becomes important for long initial free expansion times. The portion of the cloud that travels further from the trap center experiences a flatter potential that provides an insufficient momentum kick when Eqs.~(\ref{eqn: change in momentum for harmonic kick}) or~(\ref{kick strength for harmonic kick}) are used to tune the kick parameters.

\section{Canceling aberrations with multiple kicks}\label{Sec 3}

\begin{figure}
    \includegraphics[width=0.5\textwidth]{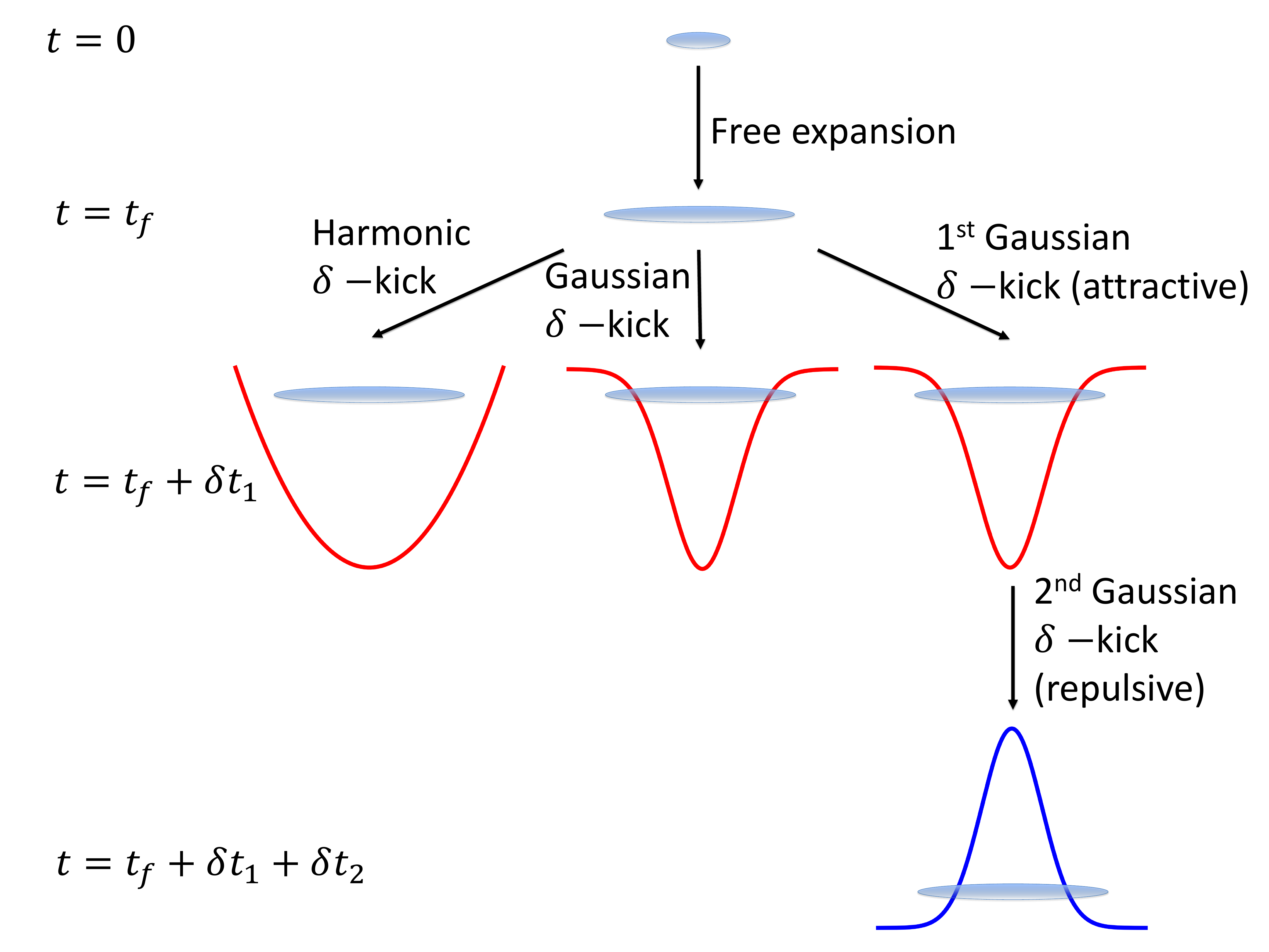}
    \caption{Schematic of the $\delta-$kick cooling protocol for an ideal harmonic kick, a single Gaussian kick, and 2 Gaussian kicks (from left to right).}
    \label{fig: cartoon of proposed protocol}
\end{figure}

\begin{figure}
    \includegraphics[width=0.5\textwidth]{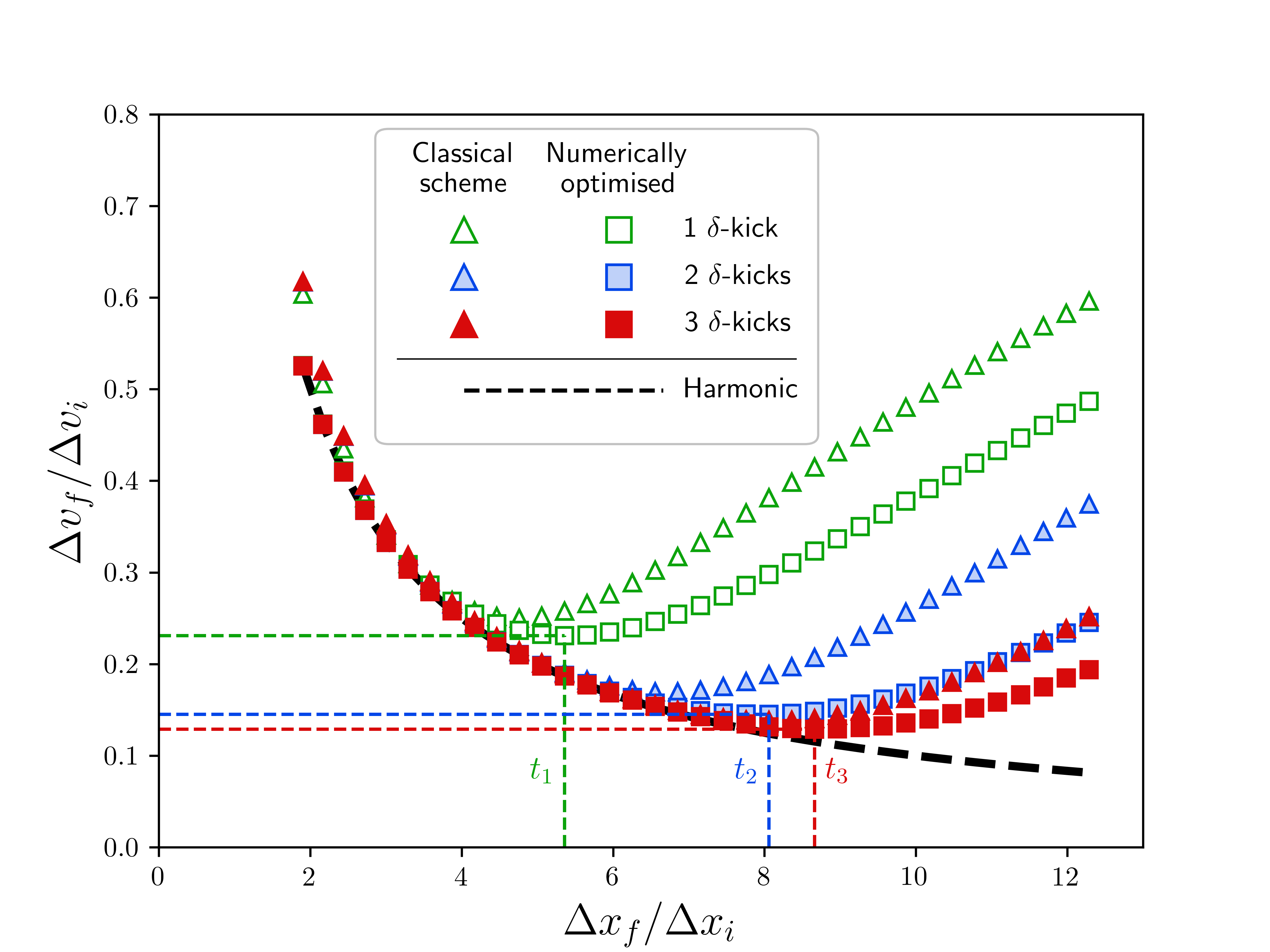}
    \caption{Comparison between final velocity width after an ideal harmonic kick (black, dashed curve), with that after 1 (green), 2 (blue), and 3 (red) Gaussian potential kicks. Triangles and squares represent data points corresponding to the classically inspired scheme and the numerically optimized case, respectively. The RMS width of the Gaussian potential is: $15\Delta x_i$ (attractive) for 1 $\delta-$kick; $15\Delta x_i$ (attractive) and $14\Delta x_i$ (repulsive) for 2 $\delta-$kicks; and $15\Delta x_i, 13\Delta x_i$ (attractive), $14\Delta x_i$ (repulsive) for the 3 $\delta-$kicks. $t_1$, $t_2$, and $t_3$ are the initial free expansion times corresponding to the best-performing singlet, doublet, and triplet lenses, respectively.}
    \label{Fig_DKC_performance_Gaussian_potential}
\end{figure}

Spherical aberrations are a well-known problem when focusing light through non-parabolic lenses. DKC using Gaussian matter-wave lenses, as described above, faces a similar problem: atoms 
which travel far from the lens center are not all brought to rest by the imperfect lens. 

We extend the idea of aberration cancellation to matter-wave lensing: compensating for the effect of non-harmonic terms in an attractive Gaussian potential kick by adding a repulsive one. We find that using multiple $\delta-$kicks increases the region over which the atoms experience an effective harmonic potential, making it possible to achieve lower kinetic temperatures with longer free expansion. A schematic of the proposed protocol is shown in Fig.~\ref{fig: cartoon of proposed protocol}. 

The use of a repulsive potential has been proposed in the context of time-optimal protocols for DKC, where a sudden trap inversion leads to faster initial expansion, which is then followed by a harmonic $\delta-$kick  \cite{DKC_David2021}. Multiple lens systems, such as a two-step trap frequency quench to improve DKC  \cite{DKC_two_step_quench} and successive magnetic lenses for 3D focusing of launched clouds \cite{Two_magnetic_lens_2006}, have also been proposed. Below, we describe an aberration cancellation setup for delta-kick cooling.

\subsection{Classically inspired scheme}\label{SubSec 3a}

We consider a compound lens system involving $N$ back-to-back matter-wave lenses. We extend the classical analysis carried out in the previous section to calculate the optimal kick strengths for this matter-wave N-tuplet. $N$ successive $\delta-$kicks $U_n(x) = U_n(1-e^{-x^2/2\sigma_n^2})$ for time $\delta t_n$ each are applied to an atom after free expansion, where $n \in \{1,2\hdots N\}$. $U_n$ is positive (negative) for an attractive (repulsive) kick. The final momentum of the atoms after the $N$ kicks is written similarly to Eq. (\ref{eqn_momentum with single Gaussian kick}), with the term 
$\left(U \delta t/\sigma^2\right) x_f e^{-x_f^2/2\sigma^2}$ replaced by $\sum_{n=1}^N \left(U_n\delta t_n / \sigma_n^2\right) x_f e^{-x_f^2/2\sigma_n^2}$. 

We define the kick strengths for the $n^{th}$ Gaussian potential $U_n(x)$ as $\kappa_n \equiv U_n\delta t_n$. The combined aim of the $N$ kicks is to reduce the effect of the non-harmonic terms in the potential. We choose the kick strengths in a manner that cancels the first $N$ terms in the Taylor expansion of the final momentum. This amounts to emulating a harmonic potential using $N$ Gaussian potentials over a region of space where terms higher than $x^{2N}$ can be ignored. This gives a set of equations, written in matrix form as

\begin{equation}  \label{eqn_matrix}
\begin{pmatrix}
\frac{1}{\sigma_1^2} & \frac{1}{\sigma_2^2} &\hdots & \frac{1}{\sigma_N^2}\\
\frac{1}{\sigma_1^4} & \frac{1}{\sigma_2^4} &\hdots & \frac{1}{\sigma_N^4} \\
\vdots &  \vdots &\hdots &\vdots \\
\frac{1}{\sigma_1^{2N}} & \frac{1}{\sigma_2^{2N}} &\hdots & \frac{1}{\sigma_N^{2N}}
\end{pmatrix}
\begin{pmatrix}
\kappa_1 \\ \kappa_2\\ \vdots \\ \kappa_N
\end{pmatrix}
 = \begin{pmatrix}
 \frac{m}{t_f}\\ 0\\ \vdots\\0
 \end{pmatrix}.
\end{equation}
Given the size of each Gaussian potential $\sigma_i$ and the free expansion time $t_f$, Eq.~(\ref{eqn_matrix}) can be used to find the optimum kick strengths for this classical scheme.

\begin{figure*}
    \includegraphics[width=1\textwidth]{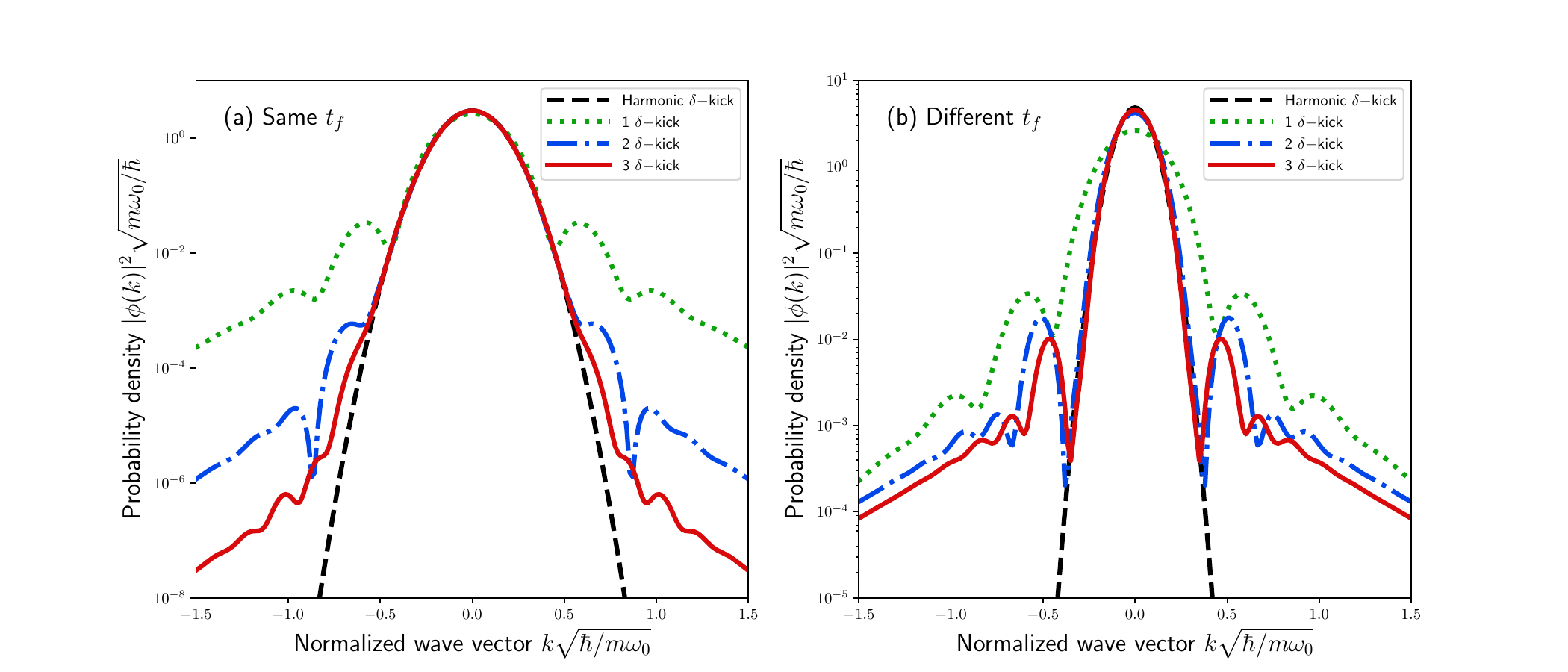}
    \caption{Momentum distributions resulting from a harmonic $\delta-$kick (black, dashed) and those resulting from 1 (green, dotted), 2 (blue, dash-dotted), and 3 (red, solid) numerically optimized Gaussian $\delta-$kicks. In (a) the same free expansion time ($t_f = t_1$) is used for each case, while in (b) $t_f=t_1, t_2, t_3, t_3$ for 1, 2, 3 Gaussian $\delta-$kicks and the harmonic $\delta-$kick respectively. The potentials have the same RMS widths as in Fig.~\ref{Fig_DKC_performance_Gaussian_potential}. }
    \label{fig: N DKC with same and different free expansion times}
\end{figure*}

\subsection{Generalized DKC inspired scheme}\label{SubSec 3B}

We note that, while the above classical analysis is intuitive and works for a non-interacting cloud of atoms in the large free expansion time limit, the approach given in \cite{DKC_David2021} is an exact quantum description of DKC with harmonic potentials for many interacting systems. For an arbitrary scale-invariant time evolution of quantum gases, an out-of-equilibrium state can be written as an equilibrium state of a new harmonic trap with trap frequency $\omega_f = \omega_0/b_f^2$, multiplied by a phase factor. Here $\omega_0$ is the frequency of the initial trap, and $b_f$ is the scaling factor for the cloud size, i.e., $\Delta x(t) = b(t) \Delta x_i$, with $b_f \equiv b(t_f) $. The time evolution of $b(t)$ for an arbitrary modulation of the trap frequency $\omega(t)$ is given by the Ermakov equation, 
\begin{equation}
    \Ddot{b} + \omega(t)^2 b = \omega_0^2/b^3.
\end{equation}
 For a cloud of $N_{a}$ atoms, the phase factor is given by $\exp{\left(i \frac{m\Dot{b}}{2\hbar b}\sum_{i=1}^{N_{a}} r_i^2\right)}$. The $\delta-$kick cooling pulse is chosen in a way that cancels this extra phase factor. For a harmonic $\delta-$kick, this gives Eq.~(\ref{kick strength for harmonic kick}) for the optimal kick parameters. In the instantaneous $\delta-$kick limit, we extend this analysis to cancel the extra phase using $N$ Gaussian potential $\delta-$kicks. This gives 
\begin{multline}
    \exp{\left(i \frac{m\Dot{b}}{2\hbar b}\sum_{i=1}^{N_{a}} r_i^2\right)} \exp{\left[\frac{-i}{\hbar} \sum_{i=1}^{N_{a}}\sum_{j=1}^N \kappa_j (1-e^{-r_i^2/2\sigma_j^2})
 \right]}\\ \approx 1.
\end{multline}
As before, we choose the kick strengths in a way that cancels the first $N$ terms in the Taylor expansion of the potentials. This gives the same final condition as Eq.~(\ref{eqn_matrix}), with the modification that the $m/t_f$ term on the right-hand side is replaced by $m \Dot{b}/b$. For free expansion of a non-interacting gas, $b(t) = \sqrt{1+\omega_0^2t^2}$. Thus, $\frac{m \Dot{b}(t_f)}{b(t_f)} = \frac{m\omega_0^2 t_f}{1+\omega_0^2 t_f^2}$. This again reduces to $m/t_f$ for large values of $t_f$, thereby recovering Eq.~(\ref{eqn_matrix}).

\subsection{
  \texorpdfstring
    {Improved compound $\delta-$kick lens performance}
    {Improved compound delta-kick lens performance}
}\label{SubSec 3C}
The approach described in Sec.~\ref{Sec 3} provides a recipe for how to design the kick strengths of the individual lenses in a compound matter-wave lens system. Here we evaluate the performance of different compound $\delta-$kicks. We compute the Wigner function (see Fig.~\ref{Wigner_fn_initial_free_harmonic_1_2_3_Gaussian} (d)-(f)), the momentum distribution (Fig.~\ref{fig: N DKC with same and different free expansion times}), and the final velocity width (Fig.~\ref{Fig_DKC_performance_Gaussian_potential}) after $N$ $\delta-$kicks. 
In the limit $\delta t_n \to 0$, the order in which the kicks are applied becomes irrelevant, and the wave function after $N$ $\delta-$kicks can be written as
\begin{equation}
        \psi(x,t_f + \delta t_1  + \dots + \delta t_N) = \left(\prod_{n=1}^N e^{-iU_n(x)\delta t_n/\hbar}\right)\psi(x,t_f).
\end{equation}
The effect of non-harmonic terms in the kick potential can be seen on the Wigner function in Fig.~\ref{Wigner_fn_initial_free_harmonic_1_2_3_Gaussian} (d), which can be corrected for by using multiple $\delta-$kicks, as in Fig.~\ref{Wigner_fn_initial_free_harmonic_1_2_3_Gaussian} (e) and (f).

Fig.~\ref{Fig_DKC_performance_Gaussian_potential} compares the cooling performance from a harmonic lens with that from singlet, doublet, and triplet Gaussian lens systems. Varying the strength of the optical potential is typically much easier experimentally than varying the width. The RMS width of the potential is therefore kept constant for our simulations. The kick strengths for each of the Gaussian $\delta-$kick cases are chosen according to the classically inspired scheme as well as through numerical optimization. 
For a harmonic $\delta-$kick, $\Delta v_f/\Delta v_i = \Delta x_i/\Delta x_f$, as in Eq.~(\ref{eqn: T_f/T_i}). More free expansion results in lower final momentum width after the $\delta-$kick without any fundamental limit on the kinetic temperature. However, for a Gaussian potential $\delta-$kick (1 $\delta$-kick in the legend), beyond a certain final size of the cloud, further free expansion does not result in a lower velocity width.
Note that the initial free expansion time that optimizes the cooling performance (marked as $t_1$, $t_2$, $t_3$ respectively in Fig.~\ref{Fig_DKC_performance_Gaussian_potential} for the singlet, doublet, and triplet matter-wave lenses) increases with the use of additional lenses. This indicates that when using compound lenses, the wavefunction can expand into a larger region that is effectively harmonic than for the singlet. 
For small initial free expansion times, the classical scheme for choosing kick parameters gives worse cooling performance than the full numerical optimization because the approximation in Eq.~(\ref{eqn_momentum with single Gaussian kick}) is more appropriate at long expansion times when $t_f\gg \Delta x_i/\Delta v_i$ and the initial cloud size is less relevant.

To further show the effect of using additional $\delta-$kicks, Fig.~\ref{fig: N DKC with same and different free expansion times} compares the full momentum distributions after a harmonic $\delta-$kick with that after $1$, $2$ and $3$ Gaussian $\delta-$kicks. The kick strengths for the Gaussian kick cases are chosen through numerical optimization to minimize the final RMS momentum width. Fig.~\ref{fig: N DKC with same and different free expansion times} (a) is plotted for the same initial free expansion time ($t_f = t_1$) for all four protocols. While the momentum distribution after a Harmonic $\delta-$kick is Gaussian throughout, the distribution after a single Gaussian $\delta-$kick shows additional side peaks and exponential tails. Using additional $\delta-$kicks lowers the amplitude of these extra features, pushing the full distribution closer to the ideal harmonic case.
This improvement is further visible in Fig.~\ref{fig: N DKC with same and different free expansion times} (b), where longer free expansion time is used for protocols with more Gaussian potential kicks. The momentum distribution becomes narrower with each additional kick. For the parameters chosen for Fig.~\ref{fig: N DKC with same and different free expansion times} (b), we observe a reduction in the kinetic temperature by a factor of $2.5$ in going from a single $\delta-$kick to 2 $\delta-$kicks and by a factor of $3.2$ in going from 1 to 3 $\delta-$kicks. The cost for this improvement is that the magnitude of required kick strengths increases with the number of kicks.

\begin{figure}
    \includegraphics[width=0.5\textwidth]{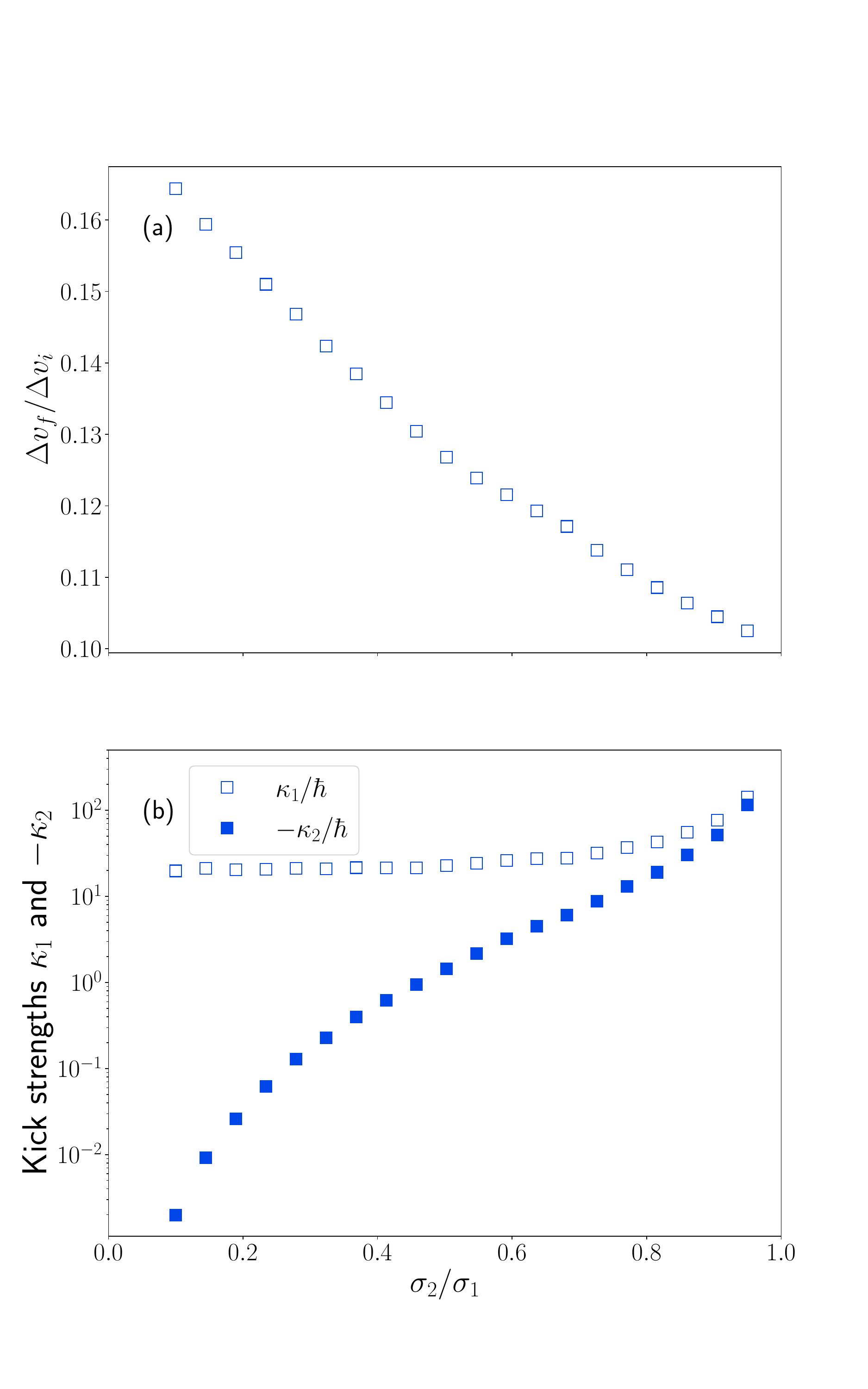}
    \caption{(a) Ratio of final and initial velocity width $\Delta v_f/\Delta v_i$, and (b) kick strengths $(\kappa_1$ and $-\kappa_2)$ for the numerically optimized doublet lens for different sizes of the repulsive $\delta-$kick beam ($\sigma_2$) (with $\sigma_1 = 15 \Delta x_i$).}
    \label{fig:two-delta-kick-performance-for-different-sigma2}
\end{figure}

Eq. (\ref{eqn_matrix}) suggests that two or more Gaussian potentials of equal widths cannot be used to form the aberration cancellation setup. To help decide on the relative size of the $\delta-$kick potentials, we repeated the above simulations for a doublet lens by varying the RMS width of the repulsive beam ($\sigma_2$). The size of the attractive beam was kept fixed at $\sigma_1 = 15\Delta x_i$. The ratio of final and initial velocity width, and the optimal kick strengths for each value of $\sigma_2/\sigma_1$ are plotted in Fig. \ref{fig:two-delta-kick-performance-for-different-sigma2} (a) and (b) respectively. We find that the cooling performance improves as the two beams get closer in size. The cost is, again, the increase in the magnitude of the required kick strengths as $\sigma_2 \to \sigma_1$.
\begin{figure}
    \includegraphics[width=0.5\textwidth]{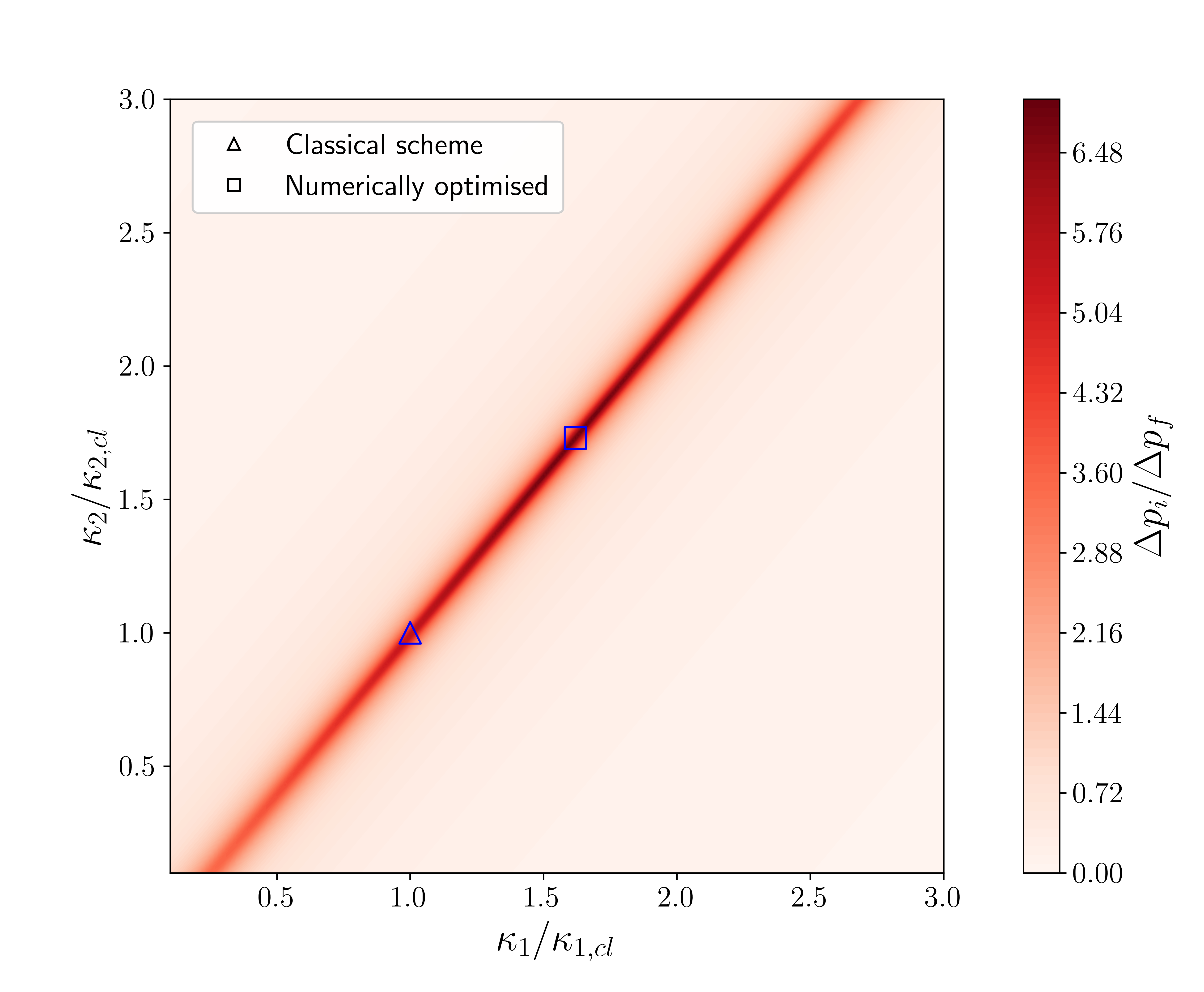}
    \caption{Ratio of initial and final momentum width after 2 $\delta-$kicks ($\sigma_1 = 15 \Delta x_i, \sigma_2 = 14\Delta x_i$, $t_f = t_2$), as a function of the kick strengths for attractive and repulsive $\delta-$kicks ($\kappa_1/\kappa_{1,cl}$ and $ \kappa_2/\kappa_{2,cl}$ respectively). The blue triangle and the square represent the kick strengths corresponding to the classically inspired ($\kappa_{1,cl}, \kappa_{2,cl}$) scheme and the numerically optimized case, respectively.}
    \label{fig:two-delta-kick-performance}
\end{figure}

Lastly, we test how sensitively the performance of the doublet lens depends on the exact value of the kick strengths chosen. To this end, we vary the kick strength around the values obtained from the classically inspired scheme (relabelled as $\kappa_{1,cl}$ and $\kappa_{2,cl}$) and plot $\Delta p_i/\Delta p_f$ as a function of $\kappa_1/\kappa_{1,cl}$ and $\kappa_2/\kappa_{2,cl}$ in Fig.~\ref{fig:two-delta-kick-performance}. The results show high sensitivity to the ratio of the kick strengths, $\kappa_1/\kappa_2$, where a few percent deviation in just one of the two kick strengths has a significant effect on the final momentum width. If $\kappa_1/\kappa_2$ is chosen correctly, however, the final momentum width is rather insensitive to common variations in the kick strengths. Furthermore, this sensitivity to the exact value of kick strengths depends on the choice of $\sigma_1$ and $\sigma_2$, where the high-performance region in Fig.~\ref{fig:two-delta-kick-performance} broadens if $\sigma_2$ decreases (keeping $\sigma_1$ constant). The choice of the relative beam sizes therefore represents a trade-off between peak cooling performance and robustness to experimental imperfections. Also, note that as in Fig.~\ref{Fig_DKC_performance_Gaussian_potential}, the numerically optimized doublet lens outperforms the doublet lens with kick parameters chosen from the classically inspired scheme.

As a possible implementation, we consider a BEC of \(^{87}\)Rb in a harmonic trap with frequency 30 Hz. The 
initial size of this cloud is \(\Delta x_i\) = 1.4 \si{\micro\meter}. We note that this assumes the cloud 
is non-interacting; for
an interacting BEC, the mean-field energy causes a larger initial cloud size in a given trap. 
Attractive kicks are applied using a Gaussian beam with \(\lambda\) = 1064 nm, and repulsive kicks are applied
using \(\lambda\) = 420 nm, blue-detuned from the 5S\(_{1/2}\) to 6P\(_{3/2}\) transition. Using a single 
attractive Gaussian kick with a width of \(\sigma_1 = 15\Delta x_i\) = 20.9 \si{\micro\meter}, 
the optimum cooling performance is achieved after an expansion time of 27.6 ms, after which the atoms 
have expanded to a final size of \(\Delta x_f = 5.36 \Delta x_i\) = 7.38 \si{\micro\meter} (Fig. \ref{Fig_DKC_performance_Gaussian_potential}), which gives a factor of 18.7 reduction in temperature.
The required kick strength is \(\kappa_1 =\) 24.2 \(\hbar\). 
If the total time allotted for the kicks is 100 \si{\micro\second}, then
achieving this kick strength requires a modest laser power of 2.22 mW. 

Adding the repulsive second kick with a width of \(\sigma_2 = 14\Delta x_i\) = 19.5 \si{\micro\meter} increases the maximum possible cooling performance. After an expansion time of 41.9 ms, the atoms have expanded to a final size of \(\Delta x_f = 8.06 \Delta x_i\) = 11.2 \si{\micro\meter}, giving a factor of 47.6 reduction in temperature and an improvement of a factor of 2.5 over the single kick. Achieving this requires kick strengths of 177.67 \(\hbar\) and -144.04 \(\hbar\) respectively for the attractive and repulsive kicks. With the same maximum time of 100 \si{\micro\second}, this requires laser powers of 16.3 mW and 14.2 mW 
for the attractive and repulsive kicks respectively.
The kicks can be applied concurrently, which 
halves the laser power requirements compared to applying the kicks consecutively. The second beam can 
counterpropagate with the first, or they may be combined using a dichroic mirror. Careful alignment is 
required to ensure that the centers of the kicking potentials coincide at the atoms to prevent unwanted 
forces on the center of mass. Translation stages may also be useful to allow for fine adjustment of the 
width of the potentials seen by the atoms. 

\section{Conclusion}\label{Sec 4}

DKC with Gaussian potentials, as it is commonly implemented in cold atom experiments, has a cooling limit set by the finite size of the potential. We have shown that using additional attractive and repulsive $\delta-$kicks can significantly improve this cooling limit. The classical description of DKC for harmonic potentials is extended to describe DKC using $N$ back-to-back Gaussian potential $\delta-$kicks. The appropriate kick parameters are obtained by canceling terms in the Taylor expansion of the final momentum. A general description of DKC in \cite{DKC_David2021} for arbitrary scale-invariant dynamics is extended similarly to determine the optimum kick parameters. The kick strengths are also determined numerically by minimizing the final RMS momentum width after the application of $N$ $\delta-$kicks. We observe that a compound matter-wave lens comprised of many individual lenses selected in this way approaches the performance of the harmonic $\delta$-kick for longer initial expansion times the more lenses are included. 

Analysis of the full momentum distribution shows extra features arising from the use of Gaussian kicking potential. For the same initial free expansion, the amplitude of these extra features is reduced with the help of additional $\delta-$kicks. More $\delta-$kicks are also used with longer initial free expansion times, resulting in lower kinetic temperatures. Specifically, for a Gaussian beam with a size 15 times the initial cloud size ($\sigma_1 = 15\Delta x_i$), we find a temperature reduction of a factor of $2.5$ when adding a second $\delta-$kick of size $\sigma_2 = 14\Delta x_i$ and a factor of $3.2$ when adding a third kick of size $\sigma_3 = 13\Delta x_i$. We have also discussed a reasonable experimental realization of this improvement using a rubidium BEC. The approach demonstrated here to reduce the effect of the anharmonic part of a Gaussian potential can be extended to potentials of other shapes. This work continues the long-standing effort to extend techniques well-developed in classical optics into the realm of quantum mechanics and matter waves. 

\section{ACKNOWLEDGMENTS}

The authors thank Adolfo del Campo, Saurabh Pandey, Deepayan Banik, and Yatharth Shivhare for helpful discussions. This work has been supported by NSERC and the Fetzer Franklin Fund.

\section*{Data Availability}
Data supporting this article is available at \url{https://github.com/harshilneeraj/delta-kick-cooling-using-multiple-Gaussian-kicks}.

\bibliography{citations}
\end{document}